\documentclass[tradiabstract]{aa} 
\usepackage{graphicx}
\usepackage{txfonts}

\usepackage{natbib}
\bibpunct{(}{)}{;}{a}{}{,} 

\begin{document}
   \title{AGN's UV and X-ray luminosities in clumpy accretion flows}

   \subtitle{}

   \author{W. Ishibashi
          \inst{1,2}
          \and
          T. J.-L. Courvoisier
          \inst{1,2}
          }

   \institute{ \textsl{INTEGRAL} Science Data Centre, ch. d'Ecogia 16, 1290 Versoix, Switzerland 
    \and Geneva Observatory, Geneva University, ch. des Maillettes 51, 1290 Sauverny, Switzerland \\
    e-mail: Wakiko.Ishibashi@unige.ch, Thierry.Courvoisier@unige.ch 
             }

   \date{Received; accepted}

 
  \abstract{
We consider the fuelling of the central massive black hole in Active Galactic Nuclei, through an inhomogeneous accretion flow.
Performing simple analytical treatments, we show that shocks between elements (clumps) forming the accretion flow may account for the UV and X-ray emission in AGNs.
In this picture, a cascade of shocks is expected, where optically thick shocks give rise to optical/UV emission, while optically thin shocks give rise to X-ray emission.
The resulting blue bump temperature is found to be quite similar in different AGNs. 
We obtain that the ratio of X-ray luminosity to UV luminosity is smaller than unity, and that this ratio is smaller in massive objects compared to less massive sources. This is in agreement with the observed $L_{X}/L_{UV}$ ratio and suggests a possible interpretation of the $\alpha_{OX}-l_{UV}$ anticorrelation.
}


   \keywords{Accretion, accretion disks - radiation mechanisms: general - galaxies: active -
   ulatraviolet: galaxies - X-rays: galaxies
               }

   \authorrunning{ W. Ishibashi \and T. J.-L. Courvoisier}
   \titlerunning{UV/X luminosities in clumpy accretion flows}

   \maketitle
%

\section{Introduction}

According to the standard paradigm of Active Galactic Nuclei (AGN) physics, black hole fuelling occurs steadily via a geometrically thin, optically thick, accretion disc. The outward transport of angular momentum is set by viscosity, generally attributed to magnetic fields and turbulent motions, characterized by the $\alpha$ parameter of the accretion disc (\citet{S_S_1973}). 
The thermal emission from the disc, peaking in the optical/UV domain, has been associated with the blue bump component (\citet{S_1978}, \citet{M_1983}). \\
 
Much less is known on the kinematics of the accreting gas at large distances, where most (up to $99.9\%$) of the angular momentum must be lost for matter to reach the $r \lesssim 100 \, R_{S}$ region where the bulk of the gravitational energy is released as radiation.
In addition, standard accretion disc models are known to face several problems when confronted with detailed observations of AGNs (\citet{C_2001}, \citet{S_et_2008}).
One of the major problems is given by the quasi-simultaneity of the UV and optical continuum variations that is not compatible with viscous time scales within discs characterized by temperature gradients (\citet{C_C_1991}, \citet{C_1991}). 
Another difficulty is the observed similarity in UV spectral properties over several orders of magnitude in luminosity, and in particular the stability of the blue bump temperature (\citet{W_et_1994}).
Moreover, a `bare' thin disc cannot account for the X-ray component of AGN emission, and therefore additional elements such as optically thin coronae, surrounding the disc and producing X-ray photons through inverse Compton processes, are required.  \\

One useful tool to study the accretion flow and radiative processes is provided by the relationship between the UV and X-ray emission components. 
In the majority of objects, observations indicate a small ratio of X-ray luminosity to blue bump luminosity. In particular, the relative importance of the hard X-ray component is found to be weaker in high luminosity AGNs (QSOs) compared to less luminous Seyfert galaxies (\citet{K_B_1999}).
This relative strength is generally quantified by the optical/UV to X-ray index $\alpha_{OX}$. 
It is now observationally confirmed that the slope of the $L_{X}(\mathrm{2 keV}) - L_{UV}(\mathrm{2500 \AA})$ relation is smaller than unity, indicating that the X-ray luminosity is smaller than the UV luminosity, and leading to the well-known $\alpha_{OX}-l_{UV}$ anticorrelation. (\citet{S_et_2005}, \citet{S_et_2006}).   
This anticorrelation implies that luminous AGNs emit less energy in the X-rays relative to the optical/UV compared to less luminous objects (\citet{J_et_2007}, \citet{K_et_2008}).
It seems that there is currently no satisfying theoretical study able to account for the small ratio of X-ray to UV luminosity, and to predict the observed $\alpha_{OX}-l_{UV}$ anticorrelation.  \\

In contrast to binary systems where the angular mometum of the accreting material is constrained by the binary geometry, accretion in AGNs proceeds in a less regular and more chaotic manner, with matter arriving from a wide range of directions. This leads to a distribution of angular momentum in the accretion flow that leads to a wide distribution of angular velocities and complex phenomenology that in turn induces numerous shocks within the accretion flow.
In view of the interest of widening the study of accretion flows and of difficulties met by disc models, we study the expected emission from the shocked material and show that several fundamental observations of AGN phenomenology can be well described by generic properties of the shocked material. 
In a previous paper, we considered an inhomogeneous flow in the form of interacting clumps of matter (\citet{C_T_2005}, hereafter C\&T 2005). We showed that shocks between clumps and the subsequent gas expansion are at the origin of the radiation and, at the same time, provide a physical mechanism of angular momentum transport.
The survival of clumps in the deep gravitational well of the central black hole was also briefly discussed. \\

The present paper is organized as follows.
The relevant features of the cascades of shocks model are briefly recalled in Sect. 2.
We then further develop the model and estimate the UV luminosity arising from the optically thick shock, which is used to determine the blue bump temperature in terms of collision parameters (Sect. 3). We then calculate the X-ray luminosities resulting from the optically thin shocks, taking into account the filling factor of the expanding clouds and considering characteristic time scales, which define different classes of objects (Sect. 4). 
We compute the luminosity ratio $L_{X}/L_{UV}$ in Sect. 5, and try to identify the different cases with different AGN classes (Sect. 6). 
We compare the model $L_{X}/L_{UV}$ ratio, as given by the relative strength of optically thin shocks compared to optically thick shocks, with observational measurements of the $\alpha_{OX}$ index (Sect. 7) and discuss resulting implications in Sect. 8.


\section{Cascades of shocks}

C\&T (2005) considered an inhomogeneous accretion flow formed by individual clumps of matter interacting with one another while accreting on to the central black hole. 
Shocks, resulting from collisions between clumps, provide the mechanism whereby gravitational energy of the ions is converted into radiative energy of the electrons. 

Model predictions were compared to observations mainly based on 3C 273 data. 
The UV lightcurve of 3C 273 has been previously described as a superposition of independent events, with total energy of $\sim$$10^{52} \, \mathrm{erg}$ and luminosity of a few $10^{44} \, \mathrm{erg/s}$ per event (\citet{P_C_W_1998}). These values correspond to the kinetic energy of clumps with masses of several $10^{33} \, \mathrm{g}$ located at a distance of $100 \, R_{S}$ ($R_{S} = \frac{2GM_{BH}}{c^{2}}$, $M_{BH}$ being the mass of the central black hole) from the centre, moving at the local free-fall velocity $v_{ff}$ ($v_{ff} = \sqrt{2GM_{BH}/R}$) on the order of $\sim$$0.1 \mathrm{c}$ . \\

At a distance of $100 \, R_{S}$, a collision between clumps moving at the local free-fall velocity $v_{ff}$ results in an optically thick shock which gives rise to optical-UV radiation. 
Assuming blackbody emission, the photospheric temperature of the expanding gas cloud can be associated with the blue bump temperature. 
In C\&T (2005) the blackbody temperature was estimated using the event luminosity derived from observations. Here, we remove the explicit dependence of the shock properties on the observational quantities and discuss the temperature self-consistently (section 3).

Closer to the centre, the expanding clouds overlap and shock again, resulting this time in optically thin shocks; hence the name of the model. Considering the equilibrium between Coulomb heating and Compton cooling, the electron temperature is estimated to be around a few hundred $\mathrm{keV}$, the optically thin shock is thus considered as the origin of the X-ray emission. 
C\&T (2005) computed the X-ray luminosity in the case of expanding clouds filling a volume comparable to the region within $100 \, R_{S}$.
In the present paper, we distinguish several cases: we consider the different locations of the optically thin shocks (depending on the volume filling factor of the expanding clouds), and estimate the X-ray luminosities taking also into account the electron radiation time.


\section{Optically thick shocks and UV emission}

Following C\&T (2005), we consider clumps of mass $M_{c} = M_{33} \cdot 10^{33} \, \mathrm{g}$ in the gravitational field of the central black hole, at a distance of $100 \, R_{S}$ moving at the local free-fall velocity $v_{ff}$. 
Expressing radial distances from the centre in terms of the Schwarzschild radius ($R = \zeta R_{S}$, $\zeta$ being a dimensionless constant), the free-fall velocity is defined by a single scaling parameter $\zeta$:
\begin{equation}
v_{ff} \left(R = \zeta R_{S}\right)  = \frac{c}{\sqrt{\zeta}} \, . 
\end{equation}
The free-fall velocity at distance of $100 \, R_{S}$ is on the order of $\sim$$0.1c$ and is given by
\begin{equation}
v_{ff} \left(R_{\zeta_{UV} R_{S}}\right) \cong 3 \cdot 10^{9}  \zeta_{UV}^{-1/2} \, \mathrm{cm/s} \, ,
\end{equation}
where $\zeta_{UV} = \zeta / 100$. \\

A collision between two such clumps results in an optically thick shock, leading to a thermalized gas cloud in rapid expansion. 
As Coulomb collisions between particles are elastic, the cloud expansion velocity is expected to be of the same order as the initial velocity: $v_{exp} \approx v_{ff}$. 
This rapid expansion is similar to a supernova explosion, for which typical explosion energies are on the order of $\sim$$10^{51} \, \mathrm{ergs}$ with velocities reaching $\sim$$10^{9} \, \mathrm{cm/s}$ (comparable to the event energy and expansion velocity, respectively). \\

Following the expansion, a fraction $\eta_{rad}$ of the kinetic energy of the colliding clumps is radiated at the photosphere. The photospheric radius is estimated by equating the photon diffusion time $t_{diff} = \frac{R^{2}}{c} \sigma_{T} n_{e}$ (where $\sigma_{T}$ is the Thomson cross section and $n_{e}$ is the electron density) and the gas expansion time $t_{exp} = \frac{R}{v_{exp}}$.
The photospheric radius is thus given by:
\begin{equation}
R_{max} \cong 3 \cdot 10^{15} \, M^{1/2}_{33}  \zeta_{UV}^{-1/4} \; \textrm{cm} \, ,
\end{equation}
and the corresponding expansion time $t_{exp}$ is:
\begin{equation}
t_{exp} \cong 10^{6} \, M^{1/2}_{33}  \zeta_{UV}^{1/4} \; \textrm{s} \, .
\end{equation} 

The resulting luminosity emitted at the photosphere is given by the fraction $\eta_{rad}$ of the kinetic energy divided by the expansion time:
\begin{equation}
L_{UV} = \frac{\eta_{rad}  M_{c} v^{2}_{ff}}{t_{exp}} 
\label{L_{UV}}
\end{equation}

The radiative efficiency $\eta_{rad}$ of the collision is estimated by analogy with a supernova explosion. In the supernova case, the initial progenitor size is important in determining the conversion efficiency: extended configurations (implying large radii $\mathrm{r} \, \sim 10^{15} \, \mathrm{cm}$) radiate more efficiently compared to compact configurations. 
Numerical models \citep{F_A_1977} predict efficiency values around $10-30 \, \%$ for extended configurations. 
In the following, we parametrize the radiative efficiency by $\eta_{rad} = \eta_{1/3} \cdot \frac{1}{3}$.

\subsection{UV luminosity}

The UV luminosity resulting from a single collision can now be expressed in terms of the collision parameters:
\begin{equation}
L_{UV} \cong 3 \cdot 10^{45} \eta_{1/3}  M^{1/2}_{33}  \zeta^{-5/4}_{UV} \; \textrm{erg/s}
\label{L_UV}
\end{equation} 

The average UV luminosity is given by the luminosity of a single event (\ref{L_UV}) multiplied by the average number of collisions, $\langle N_{c} \rangle$.  This number may be estimated by considering mass conservation (which implies that clumps accretion and destruction time scales are on the same order and are given by the expansion time) and taking into account only collisions per pair: $\langle N_{c} \rangle = \langle N \rangle/2$ where $\langle N \rangle = \frac{\dot{M}}{M_{c}} t_{exp} \cong  0.8 \, M_{33}^{-1/2} \zeta_{UV}^{1/4} \dot{M}$ is the number of clumps arriving in the region, on average. The mass accretion rate, $\dot{M}$, can be expressed in units of ten solar masses per year ($\frac{\dot{M}}{10 \, M_{\odot}/\mathrm{yr}}$), for a luminous quasar. 
Combining the above expressions, we get:
\begin{equation}
\langle L_{UV} \rangle = L_{UV} \cdot \left\langle N_{c} \right\rangle \cong  1.2 \cdot 10^{45} \eta_{1/3} \zeta^{-1}_{UV} \left(\frac{\dot{M}}{10 \, M_{\odot}/\textrm{yr}}\right) \; \textrm{erg/s} \, .
\label{L_UV_II} 
\end{equation}

In the case of less massive AGNs, as Seyfert galaxies, a more typical value of the accretion rate is on the order of $1 \, M_{\odot}/\textrm{yr}$, the average UV luminosity is then
\begin{equation}
\langle L_{UV} \rangle  \cong  1.2 \cdot 10^{44} \eta_{1/3} \zeta^{-1}_{UV} \left(\frac{\dot{M}}{1 \, M_{\odot}/\textrm{yr}}\right) \; \textrm{erg/s} \, .
\label{L_UV_I} 
\end{equation}

We note that the average UV luminosity scales linearly with mass accretion rate, but does not depend explicitly on the black hole mass. This explains that a large range of Eddington ratios can be accounted for by the model.

\subsection{Blue bump temperature}

Since the shock is optically thick, the resulting luminosity emerges as blackbody radiation from the photosphere. 
The photospheric temperature is estimated assuming blackbody emission:
\begin{equation}
T = \left(\frac{L_{UV}}{4 \pi \sigma R_{max}^{2}}\right)^{1/4} \; \textrm{K} \, ,
\end{equation} 
where $\sigma$ is the Stefan-Boltzmann constant, $R_{max}$ the photospheric radius, and $L_{UV}$ the luminosity of the optically thick shock given in (\ref{L_UV}). 
We estimate the blackbody temperature using the UV luminosity of a single event,
to find: 
\begin{equation}
T \cong 2.6 \cdot 10^{4} \; \eta_{1/3}^{1/4} M_{33}^{-1/8} \zeta_{UV}^{-3/16} \; \textrm{K} \, .
\label{T}
\end{equation}

The value of the blackbody temperature of a few $10^{4} \, \mathrm{K}$, given in ($\ref{T}$), corresponds to the typically observed value of the blue bump temperature, suggesting that optically thick shocks may be at the origin of the optical-UV emission in AGNs. 
Contrary to standard accretion discs, the above expression for the blue bump temperature has no explicit dependence on black hole mass: it is determined only by collision parameters. 
In addition, collision parameters intervene in expression ($\ref{T}$) with low powers, indicating a weak dependence of the blue bump temperature on these quantities. \\ 

C\&T (2005) analysed the subsequent temperature evolution and showed that the time delay increases in agreement with the observed lags between light curves of different wavelengths.


\section{Optically thin shocks and X-ray emission}

Following the optically thick shocks, the resulting gas envelopes expand. This expansion leads to interactions of the material originating in different regions, giving rise to optically thin shocks.
The expanding regions fill a volume $\sim$$R_{max}^{3}$.
The location of the second shock is mainly determined by the volume filling factor of the post-expansion configuration in the region within $100 \, R_{S}$:
\begin{equation}
\epsilon = \left( \frac{R_{max}}{R_{\zeta_{UV} R_{S}}} \right)^{3} \cong 10^{-3} \, M_{33}^{3/2} \zeta_{UV}^{-15/4} \left( \frac{M_{BH}}{10^{9} M_{\odot}} \right)^{-3}
\end{equation}

We analyse two distinct classes of objects, Class S and Class Q, according to the relative importance of the volume filling factor $\epsilon$. 
Class S objects are characterized by a large filling factor with $R_{max} \sim 100 \, R_{S}$, a condition met for relatively small central black holes ($M_{BH} \lesssim 10^{8} M_{\odot}$), for which $\epsilon$$\sim$1. 
Class Q describes the case of a small filling factor of the post-shock configuration, associated with massive AGNs ($M_{BH} \gtrsim 10^{9} M_{\odot}$), for which $\epsilon$$\sim$$10^{-3}$. 

We shall adopt the following parametrizations for the central mass and the accretion rate: in Class S, the black hole mass is expressed in units of $M_{BH} = M_{8} \cdot 10^{8} \, M_{\odot}$ and the accretion rate in units of $\dot{M} = \dot{M}_{0} \cdot 1 \, M_{\odot}/\mathrm{yr}$ ; in Class Q, the black hole mass is expressed in units of $M_{BH} = M_{9} \cdot 10^{9} \, M_{\odot}$ and the accretion rate in units of $\dot{M} = \dot{M}_{1} \cdot10 \, M_{\odot}/\mathrm{yr}$. We note that the ratio of mass accretion rate to central mass, $\dot{M}/M_{BH}$, is a variable parameter which allows us to consider a range of Eddington ratios.

\subsection{Electron energy}
\label{e_en}

Electrons are heated by Coulomb collisions with hot protons and cooled through Compton emission. The electron temperature is determined by the equilibrium between Coulomb heating and Compton cooling.

Assuming the relative speed $v_{rel}$ of the expanding clouds to be about the local expansion velocity ($\sim$ 0.1 c), we estimate that the temperature involved in this second shock is on the order of 1 MeV. The proton kinetic energy is therefore parametrized by $E_{p} = E_{p,MeV} \cdot 1 MeV$.\\

The average electron energy is estimated assuming equilibrium between Coulomb heating and Compton cooling. The Compton cooling rate in the non-relativistic limit is given by
\begin{equation}
L_{Compton} = \frac{8 \sigma_{T}}{3 m_{e} c} \cdot u_{ph} \cdot E_{e} \, ,
\label{L_{Compton2}}
\end{equation}
$u_{ph}$ being the photon energy density and $E_{e}$ the electron energy.
The photon energy density is given by the average UV luminosity of the optically thick shocks contained in the region within $100 \, R_{S}$
\begin{equation}
u_{ph} = \frac{\left\langle L_{UV} \right\rangle}{4 \pi \left(\zeta_{UV} R_{S}\right)^{2} c} \, .
\label{u_ph}
\end{equation} 

The heating rate of the electrons through Coulomb collisions, $L_{Coulomb}$, is calculated as in C\&T (2005), using however the electron number density estimated as
\begin{equation}
n = \frac{\dot{N}}{4 \pi \left(\zeta R_{S}\right)^{2}  v_{ff}\left(\zeta R_{S}\right)} \, ,
\label{n}
\end{equation}
where $\dot{N} = \frac{f \dot{M}}{m_{p}} $ is the number accretion rate, with $f$ the fraction of accreted matter contributing to the optically thin shock, and $v_{ff}(\zeta R_{S})$ the free-fall velocity at a distance of $\zeta R_{S}$. 
Using the photon energy density from $(\ref{u_ph})$ and the electron number density from $(\ref{n})$, the equilibrium condition $L_{Compton} = L_{Coulomb}$ leads to the average electron energy. 
In the following, we analyse separately the case of large and small filling factors.

\subsubsection{Class S: large filling factors}

As the post-shock configuration fills a volume comparable to the region within $100 \, R_{S}$ ($\epsilon$$\sim$1), expanding envelopes resulting from the first optically thick shocks, rapidly overlap and optically thin shocks already take place at $\sim$$100 \, R_{S}$. 
The electron number density ($\ref{n}$) is estimated within a region of $100 \, R_{S}$, with $f = f_{1} \cdot 1$. Using this electron number density and the photon energy density from ($\ref{u_ph}$), the equilibrium condition $L_{Compton} = L_{Coulomb}$ gives:
\begin{equation}
\frac{E_{e}}{m_{e}c^{2}} \cong 0.6 \; f_{1}^{2/7} \eta_{1/3}^{-2/7}  \zeta_{UV}^{3/7}  E_{p,MeV}^{4/7} \, .
\label{E_e_I}
\end{equation}
 
The average electron energy is on the order of $\sim$300 keV and does not depend explicitly on the black hole mass and accretion rate; moreover the exponents appearing in expression ($\ref{E_e_I}$) are small, indicating that collision parameters do not strongly influence the average electron energy. 
The Compton luminosity emitted by a single non-relativistic electron can be estimated inserting the electron energy  (\ref{E_e_I}) in equation ($\ref{L_{Compton2}}$):
\begin{equation}
L_{Compton} \cong 1.2 \cdot 10^{-12} \; f_{1}^{2/7} \eta_{1/3}^{5/7}  \zeta_{UV}^{-18/7} E_{p,MeV}^{4/7} \dot{M}_{0} M_{8}^{-2} \quad \textrm{erg/s} \, .
\label{Compton_lum_I}
\end{equation}

\subsubsection{Class Q: small filling factors}

As the volume filling factor is small (i.e. $\epsilon$$\ll$1), expanding envelopes have to travel a greater distance towards the centre before overlapping and interacting with one another. The optically thin shocks occur therefore in the central regions, the location of this second shock being parametrized by $\zeta_{X} = \zeta/10$. The electron number density as given in equation ($\ref{n}$) is now estimated at a distance of $\zeta_{X} \, R_{S}$ with its corresponding free-fall velocity. 
Here, we consider that only a fraction of the total accreted matter contributes to the second shock, parametrizing it as: $f = f_{1/2} \cdot \frac{1}{2}$.  
We roughly estimate that half of the accreted matter is ejected, following the first shock, giving perhaps rise to an outflow (we discuss the possibility of an outflow in Sect. 8). 
Considering the equilibrium between Coulomb heating and Compton cooling, we obtain the average electron energy
\begin{equation}
\frac{E_{e}}{m_{e}c^{2}} \cong 1.4 \; f_{1/2}^{2/7} \eta_{1/3}^{-2/7}  \zeta_{X}^{-3/7} \zeta_{UV}^{6/7} E_{p,MeV}^{4/7} \, , 
\label{E_e_II}
\end{equation} 
and its corresponding single electron Compton luminosity
\begin{equation}
L_{Compton} \cong 2.5 \cdot 10^{-13} \; f_{1/2}^{2/7} \eta_{1/3}^{5/7}  \zeta_{UV}^{-18/7} E_{p,MeV}^{4/7} \dot{M}_{1} M_{9}^{-2} \quad \textrm{erg/s} \, .
\label{Compton_lum_II}
\end{equation}

\subsection{Time scales and X-ray luminosity}

The Compton cooling of the hot electrons discussed in $\ref{e_en}$ gives rise to X-ray emission
as seen from the values of the average electron energy.
To estimate the emitted X-ray luminosity, we need to compare the relative importance of radiation and accretion timescales, determining whether electrons have enough time for radiating all their energy before disappearing in the black hole. 
The Compton cooling time is defined as:
\begin{equation}
t_{Compton} = \frac{E_{e}}{L_{Compton}} \, ,
\label{Compton_time}
\end{equation} 
where $E_{e}$ is the average electron energy and $L_{Compton}$ its corresponding luminosity. 
This quantity gives the characteristic time scale for the cooling of the electrons by Compton emission. 
The dynamical time $t_{dyn}$ is on the order of the free-fall time $t_{ff}$ at a given distance from the centre
\begin{equation}
t_{dyn} \sim t_{ff} = \sqrt{\frac{R^{3}}{2 G M_{BH}}} \, .
\label{dynamical_time}
\end{equation} 
 
We analyse two distinct cases, Case A and Case B, depending on whether the ratio of the Compton time over the dynamical time $\frac{t_{Compton}}{t_{dyn}}$ is smaller or greater than unity. 
This time scale condition translates into a condition on the ratio of accretion rate to central mass, $\dot{M}/M_{BH}$, a measure of the Eddington ratio $L/L_{Edd}$, as:
\begin{equation}
\frac{L}{L_{Edd}} \propto \frac{\eta \dot{M} c^{2}}{M_{BH}} \propto \frac{\dot{M}}{M_{BH}} \, ,
\label{Mdot/M}
\end{equation}
where $\eta$ is the conversion efficiency.

\subsubsection{Class S: large filling factors}

In the case of large filling factors, the Compton time is calculated from equation ($\ref{Compton_time}$) using expressions ($\ref{E_e_I}$), and ($\ref{Compton_lum_I}$); while the dynamical time ($\ref{dynamical_time}$) is taken at $\sim$$\zeta_{UV} \, R_{S}$.
Calculating the ratio of these two time scales, $\frac{t_{Compton}}{t_{dyn}}$, we see that the Compton time is shorter than the dynamical time provided that the ratio of mass accretion rate to central mass exceeds a critical value:
\begin{equation}
\frac{t_{Compton}}{t_{dyn}} < 1 \;  \Leftrightarrow \;  \left(\frac{\dot{M}}{1 \, M_{\odot}/\mathrm{yr}}\right) /  \left(\frac{M_{BH}}{10^{8} \, M_{\odot}}\right) > 0.4 \; \eta_{1/3}^{-1}  \zeta_{UV}^{3/2}
\label{timescale_I}
\end{equation}

The X-ray luminosity should then be calculated according to the relevant time scale. 
We consider separately two cases: Case A defined by the condition that the Compton time is larger than the dynamical time, and Case B in which the Compton time is shorter than the dynamical time. This distinction can also be expressed in terms of the Eddington ratio through relation ($\ref{Mdot/M}$).

\subsubsection*{Case A : $t_{Compton} > t_{dyn} \;  \Leftrightarrow \;  \left(\frac{\dot{M}}{1 \, M_{\odot}/\mathrm{yr}}\right) /  \left(\frac{M_{BH}}{10^{8} \, M_{\odot}}\right) < 0.4 \, \eta_{1/3}^{-1}  \zeta_{UV}^{3/2}$}

If the cooling time is long compared to the dynamical time  (i.e. $t_{Compton} > t_{dyn}$), the  average X-ray luminosity is given by the Compton luminosity emitted by a single non-relativistic electron multiplied by the average number of electrons present in the region
\begin{equation}
\langle L_{X} \rangle \; \sim \; L_{Compton} \cdot \langle N_{e} \rangle \, .
\label{L_X_IA_gen} 
\end{equation}

The average number of electrons $\langle N_{e} \rangle$ may be considered as a constant and can be estimated assuming an equilibrium situation in a spherical shell. At equilibrium, the number of incoming accreted electrons should be equal to the number of infalling advected electrons: $f \dot{M}_{acc} = \langle N_{e} \rangle / {t_{dyn}} \cdot m_{p}$, with the incoming rate set by the external accretion rate and the infall rate estimated by assuming that electrons fall towards the black hole on a dynamical time scale. 
From  (\ref{L_X_IA_gen}), the average X-ray luminosity of the optically thin shocks is on the order of
\begin{equation}
\left\langle L_{X} \right\rangle \; \cong \; 4.8 \cdot 10^{43} \; f_{1}^{9/7} \eta_{1/3}^{-2/7} \zeta_{UV}^{-15/14} E_{p,MeV}^{4/7} \dot{M}_{0}^{2} M_{8}^{-1} \quad \textrm{erg/s} \, ,
\label{L_X_IA}
\end{equation}
with explicit dependences on accretion rate and black hole mass.

\subsubsection*{Case B : $t_{Compton} < t_{dyn} \;  \Leftrightarrow \;  \left(\frac{\dot{M}}{1 \, M_{\odot}/\mathrm{yr}}\right) /  \left(\frac{M_{BH}}{10^{8} \, M_{\odot}}\right) > 0.4 \, \eta_{1/3}^{-1}  \zeta_{UV}^{3/2}$}

As the Compton time is short compared to the dynamical time  (i.e. $t_{Compton} < t_{dyn}$), the cooling process is very efficient and all the electron energy can be radiated away. 
The X-ray luminosity is given by the total energy of a single non-relativistic electron mutiplied by the number of electrons per unit time arriving in the region
\begin{equation}
\langle L_{X} \rangle \; \sim \;  E_{e} \cdot \langle \dot{N} \rangle  \, ,
\label{L_X_IB}
\end{equation} 
where $\langle \dot{N} \rangle$ is given by the mass accretion rate.
From equation  (\ref{L_X_IB}), we get:
\begin{equation}
\left\langle L_{X} \right\rangle \; \cong 2.0 \cdot 10^{43} \; f_{1}^{9/7} \eta_{1/3}^{-2/7}  \zeta_{UV}^{3/7} E_{p,MeV}^{4/7} \dot{M}_{0}  \quad \textrm{erg/s} \, .
\label{L_{X}_IB}
\end{equation}

In this case, the average X-ray luminosity scales linearly with the mass accretion rate, but is independent of the central mass.

\subsubsection{Class Q: small filling factors}

We perform identical calculations as for the Class S case, but taking into account the different location of the optically thin shocks.
The time scale condition ($\ref{timescale_I}$) is now modified as
\begin{equation}
\frac{t_{Compton}}{t_{dyn}} < 1 \; \Leftrightarrow \;  \left(\frac{\dot{M}}{10 \, M_{\odot}/\mathrm{yr}}\right) /  \left(\frac{M_{BH}}{10^{9} \, M_{\odot}}\right) > 14 \; \eta_{1/3}^{-1}  \zeta_{X}^{-3/2} \zeta_{UV}^{3}
\label{timescale_II}
\end{equation}

Applying analogous arguments as in the previous case, the average X-ray luminosities are respectively given by 

\subsubsection*{Case A: \\
$t_{Compton} > t_{dyn} \;  \Leftrightarrow \;  \left(\frac{\dot{M}}{10 \, M_{\odot}/\mathrm{yr}}\right) /  \left(\frac{M_{BH}}{10^{9} \, M_{\odot}}\right) < 14 \, \eta_{1/3}^{-1} \zeta_{X}^{-3/2}  \zeta_{UV}^{3}$} 

\begin{equation}
\left\langle L_{X} \right\rangle \; \cong \; 1.6 \cdot 10^{43} \, f_{1/2}^{9/7} \eta_{1/3}^{-2/7}  \zeta_{X}^{15/14} \zeta_{UV}^{-8/7} E_{p,MeV}^{4/7} \dot{M}_{1}^{2} M_{9}^{-1} \;\textrm{erg/s} 
\label{L_X_IIA}
\end{equation}

\subsubsection*{Case B: \\ 
$t_{Compton} < t_{dyn} \;  \Leftrightarrow \;  \left(\frac{\dot{M}}{10 \, M_{\odot}/\mathrm{yr}}\right) /  \left(\frac{M_{BH}}{10^{9} \, M_{\odot}}\right) > 14 \, \eta_{1/3}^{-1} \zeta_{X}^{-3/2}  \zeta_{UV}^{3}$} 

\begin{equation}
\left\langle L_{X} \right\rangle \; \cong 2.2 \cdot 10^{44} \, f_{1/2}^{9/7} \eta_{1/3}^{-2/7}  \zeta_{X}^{3/7} \zeta_{UV}^{6/7} E_{p,MeV}^{4/7} \dot{M}_{1}  \; \textrm{erg/s}
\label{L_X_IIB}
\end{equation} 

In Class Q, the time scale condition ($\ref{timescale_II}$) implies that massive sources should mostly fall into Case A, unless the accretion rate is extremely high. 
Indeed, a source of $10^{9} \, M_{\odot}$, accreting at the Eddington limit would have $\left(\frac{\dot{M}}{10 \, M_{\odot}/\mathrm{yr}}\right) /  \left(\frac{M_{BH}}{10^{9} \, M_{\odot}}\right) \sim 2 \, \eta_{1/3}^{-1} \zeta_{X}^{-3/2}  \zeta_{UV}^{3}$, much less than the limit $14 \, \eta_{1/3}^{-1} \zeta_{X}^{-3/2}  \zeta_{UV}^{3}$.
In the following, we therefore assume that massive objects always belong to Case A.


\section{$\frac{L_{X}}{L_{UV}}$ ratio}

The relative importance of X-ray and UV contributions to the bolometric luminosity varies in different classes of AGNs. 
In the framework of the model presented here, the relative importance of these two emission components is determined by the respective importance of optically thin shocks compared to optically thick shocks. 
This is quantified by the luminosity ratio $L_{X}/L_{UV}$, calculated as the ratio of average X-ray luminosity to average UV luminosity. 
In the following, we discuss the luminosity ratio $L_{X}/L_{UV}$ for the different cases presented in the previous section, analysing in particular its dependence on central mass and accretion rate.

\subsection{Class S: large filling factors}

The $L_{X}/L_{UV}$ ratio is calculated separately for Case A and Case B, according to the relevant time scale. From equation ($\ref{L_UV_I}$) for the UV luminosity, and expressions ($\ref{L_X_IA}$), ($\ref{L_{X}_IB}$) for the X-ray luminosities, we obtain
\begin{equation}
\textrm{Case A:} \quad
\frac{\left\langle L_{X} \right\rangle}{\left\langle L_{UV} \right\rangle} \, \cong \, 0.40  \, f_{1}^{9/7} \eta_{1/3}^{-2/7}  \zeta_{UV}^{-1/14} E_{p,MeV}^{4/7} \dot{M}_{0} M_{8}^{-1} 
\end{equation} 
\begin{equation}
\textrm{Case B:} \quad
\frac{\left\langle L_{X} \right\rangle}{\left\langle L_{UV} \right\rangle} \, \cong \, 0.17  \, f_{1}^{9/7} \eta_{1/3}^{-9/7}  \zeta_{UV}^{10/7} E_{p,MeV}^{4/7}  
\label{ratio_IB}
\end{equation}
where the black hole mass is expressed in units of $10^{8} \, M_{\odot}$, and the accretion rate in units of $1 \, M_{\odot}/\mathrm{yr}$.

We observe that the ratio of X-ray luminosity to UV luminosity is smaller than unity.
This result depends on several parameters, such as the fraction of accreted matter contributing to the optically thin shock and the radiative efficiency of the collision. Nevertheless, varying these parameters within the allowed range ($f \leq 1$ and $\eta_{rad} \sim 10-30 \%$) always leads to $L_{X}/L_{UV}$ ratios smaller than unity.

In Case A, the $L_{X}/L_{UV}$ ratio is proportional to the mass accretion rate and to the inverse of the central mass; while in Case B, it is independent of both parameters and stabilizes at a constant value when the $\dot{M}/M_{BH}$ ratio exceeds a critical value (see Fig. 1).

\subsection{Class Q: small filling factors}

In the case of small filling factors, the luminosity ratio $L_{X}/L_{UV}$ is calculated using equation ($\ref{L_UV_II}$) for the UV luminosity and expression ($\ref{L_X_IIA}$) for the X-ray luminosity
\begin{equation}
\textrm{Case A:} \quad
\frac{\left\langle L_{X} \right\rangle}{\left\langle L_{UV} \right\rangle} \, \cong \, 0.01  \, f_{1/2}^{9/7} \eta_{1/3}^{-2/7}  \zeta_{X}^{15/14} \zeta_{UV}^{-8/7} E_{p,MeV}^{4/7} \dot{M}_{1} M_{9}^{-1} 
\end{equation} 
where the black hole mass is expressed in units of $10^{9} \, M_{\odot}$, and the accretion rate in units of $10 \, M_{\odot}/\mathrm{yr}$. 

As in the previous case, the luminosity ratio is smaller than unity, with $L_{X}/L_{UV}$ being proportional to $\dot{M}/M_{BH}$.
Here, the $L_{X}/L_{UV} - \dot{M}/M_{BH}$ relation is much weaker (with a slope of $\sim 0.01$) than in the case of large filling factors. 
We note that the $L_{X}/L_{UV}$ ratio is one order of magnitude smaller in massive objects compared to less massive sources. \\ 

Summarising, our model gives X-ray to UV ratios always smaller than unity, with a luminosity ratio roughly in the range $0.01 \lesssim L_{X}/L_{UV} \lesssim 0.8$ depending on the value of the different parameters.
Another result is that massive AGNs emit less of their radiative energy in the X-rays relative to the optical/UV compared to less massive objects. 
The model luminosity ratios of the different cases discussed are given in table ($\ref{table_model}$).

\begin{table}[h!tpb]
\caption{Model $L_{X}/L_{UV}$ ratios}        
\label{table_model}      
\centering                          
\begin{tabular}{l c c}        
\hline\hline                 
         Case       & Class S &  Class Q \\    
\hline
         A & 0.40 $\dot{M}_{0}/M_{8}$  & 0.01 $\dot{M}_{1}/M_{9}$ \\         
         B & 0.17 & -\\                     
\hline                               
\end{tabular}
\end{table}


\section{Identification with different AGN classes}
\label{Identification}

The relative importance of the filling factor defines two classes of objects, their distinction being mainly determined by the mass of the central black hole, which directly sets the size of the Schwarzschild radius. 
Each class is further subdivided into two cases (Case A and Case B) separated by the time scale condition, which translates into a condition on the ratio of accretion rate to central mass.
The distinction between different classes is thus primarily determined by two of the black hole fundamental parameters: central mass and accretion rate. \\

Class Q objects describe massive objects for which the Eddington luminosity is $L_{Edd} \cong 10^{47} M_{9} \, \textrm{erg/s}$ and should be identified with massive luminous quasars, while Class S dealing with smaller black holes should be associated with less luminous sources, as Seyfert galaxies.
According to our model, a Seyfert galaxy of given central mass, would have a luminosity ratio given by $L_{X}/L_{UV} \sim 0.40 \, \dot{M}/M_{BH}$ for low accretion rates and $L_{X}/L_{UV} \sim 0.17$ for higher accretion rates (Case A and Case B respectively). 
On the other hand, massive QSOs should have luminosity ratios one order of magnitude smaller, with a predicted value given by $L_{X}/L_{UV} \sim 0.01 \, \dot{M}/M_{BH}$ (Case A); as previously mentioned, Case B is probably not realised and no such class of objects should be observed.


\section{Comparison with observations}

\subsection{$\alpha_{OX} - \frac{L_{X}}{L_{UV}}$ relation}

X-ray and UV emissions contribute quite differently to the overall luminosity in different classes of AGNs. 
The relative importance of the UV and X-ray components is generally quantified by the optical/UV to X-ray index $\alpha_{OX}$:
\begin{equation}
\alpha_{OX} = \frac{\textrm{log} \left[f (\nu_{X})/f (\nu_{UV})\right]}{\textrm{log} (\nu_{X}/\nu_{UV})}
\, \cong \, 0.384 \, \textrm{log} \frac{f_{2 keV}}{f_{2500 \AA}}
\label{alpha}
\end{equation}
defined as the slope of a hypothetical power law relating the two emission regions in the object's rest frame. 
The monochromatic rest frame fluxes are measured at 2 keV and 2500 $\AA$ for the X-ray and UV components, respectively.

In order to compare our model results with observations, we need to convert the $\alpha_{OX}$ indices used in the literature into luminosity ratios $L_{X}/L_{UV}$. 
Following \citet{A_C_2000}, we apply a correction factor $K$  (which takes into account the broader X-ray component) to roughly estimate the broadband X/UV ratio. 
The ratio of the integrated fluxes is thus approximated by
\begin{equation}
\frac{\int^{\nu_{2}}_{\nu_{1}} f_{X} d \nu }{\int^{\nu_{2}^{'}}_{\nu_{1}^{'}} f_{UV} d \nu^{'}}
\; \cong \; K \cdot \frac{\nu_{X}}{\nu_{UV}} \frac{f_{X}}{f_{UV}} \; \cong \; \frac{L_{X}}{L_{UV}} \, ,
\label{integrated_flux} 
\end{equation}
where the correction factor $K$ is determined from observations. 
\citet{A_C_2000} assume a value of $K = 4$  
(with the UV and X-ray fluxes measured at 1375 $\AA$ and 2 keV, respectively) 
for the \citet{G_et_1996} sample of Seyfert 1 galaxies, which leads to a luminosity ratio on the order of $L_{X}/L_{UV} \sim 0.3$. 

Based on the 3C 273 spectrum, and considering the 2-20 keV band for the X-rays and 3000-1300 $\AA$ band for the UV component, we verify that the correction factor is indeed in a similar range: $K \sim$ 3-4. In the following we assume $K = 3$ with the UV and X-ray components taken at 2500 $\AA$ and 2 keV, respectively. 
With this assumption, we can relate the $\alpha_{OX}$ index with the $L_{X}/L_{UV}$ luminosity ratio. From equations  (\ref{alpha}) and  (\ref{integrated_flux}), we obtain the expression relating the two quantities:
\begin{equation}
\frac{L_{X}}{L_{UV}} \; \cong \; K \cdot \frac{\nu_{X}}{\nu_{UV}} \cdot 10^{\frac{\alpha_{OX}}{0.384}} \, .
\label{L_ratio}
\end{equation}

\subsection{Observations}

The measurement of the $\alpha_{OX}$ index and its relationship with source parameters such as luminosity and redshift, has been the subject of many recent observational efforts (\citet{S_et_2005}, \citet{S_et_2006}, \citet{J_et_2007}, \citet{K_et_2008}).
The main result of these studies is the now well-established $\alpha_{OX}-l_{UV}$ anticorrelation, where $l_{UV} = \mathrm{log} L_{2500 \AA}$ is the logarithm of the UV monochromatic luminosity  (expressed in units of $\mathrm{erg s^{-1} Hz^{-1}}$).

\subsubsection{Samples}

As we are interested in emission mechanisms directly associated with the accretion phenomenon, samples used in studying the $\alpha_{OX}-l_{UV}$ index generally exclude radio-loud AGNs and broad absorption line  (BAL) objects. But it should be noted that the removal of these peculiar sources is not always straightforward. \\

\citet{S_et_2005} analysed a sample of 228 optically selected AGNs spanning a redshift range of $z = 0.01-6.3$ formed by a main sample of 155 objects selected from the Sloan Digital Sky Survey  (SDSS), with 36 additional high-redshift luminous AGNs and 37 low-redshift Seyfert 1 galaxies. 
\citet{S_et_2006} extended the \citet{S_et_2005} work, including 52 moderate-luminosity AGNs selected from the COMBO survey and 46 low-redshift luminous AGNs from the Bright Quasar Survey  (BQS). 
A representative sample of 59 of the most optically luminous quasars in the Universe in the redshift range $z = 1.5-4.5$ was studied by \citet{J_et_2007}. 
More recently, \citet{K_et_2008} performed the largest study to date of the X-ray properties of radio-quiet quasars, analysing a sample of 318 RQQs spanning a broad range in black hole mass  ($10^{6} \lesssim M_{BH}/M_{\odot} \lesssim 10^{10}$).

\subsubsection{Observational results and confrontation with model predictions}

A significant correlation between X-ray and UV emissions, described as $l_{X} \propto l_{UV}^{\beta}$, has been observed. The slope of this $l_{X}-l_{UV}$ relation is found to be inconsistent with unity, and is better characterized by $\beta \sim 0.7$. 
This implies that the ratio between the 2 keV and 2500 $\AA$ monochromatic luminosities varies with rest-frame UV luminosity. Equivalently, a clear trend indicating a significant anticorrelation between $\alpha_{OX}$ and monochromatic UV luminosity is seen when plotting the $\alpha_{OX}$ index as a function of $l_{UV}$ for a sample of optically selected AGNs  (\citet{S_et_2005}, \citet{S_et_2006}).
In optically selected samples, $\alpha_{OX}$ indices lie typically in the range $-1.7 \lesssim \alpha_{OX} \lesssim -1.3$. Specifically, \citet{S_et_2005} measured a median $\left\langle \alpha_{OX} \right\rangle = -1.51$ for the main SDSS sample, $\left\langle \alpha_{OX} \right\rangle = -1.72$ for the high-redshift sample, and $\left\langle \alpha_{OX} \right\rangle = -1.34$ for the Seyfert 1 sample. These observations indicate that lower luminosiy AGNs have flatter $\alpha_{OX}$ indices compared to higher luminosity objects.

This overall trend is further reinforced by the study of \citet{J_et_2007} who analysed a sample of the most luminous QSOs, obtaining steeper slopes with a mean value of $\left\langle \alpha_{OX} \right\rangle = -1.80$.  
The relationship between the $\alpha_{OX}$ index and the black hole mass has been recently confirmed by \citet{K_et_2008}, who studied the direct dependence of $\alpha_{OX}$ on $M_{BH}$, finding that radio-quiet quasars become more X-ray quiet as the central mass increases. \\

Table 2 summarizes all the mean $\left\langle \alpha_{OX} \right\rangle$ values from the above quoted papers along with their corresponding $\left\langle L_{X}/L_{UV} \right\rangle_{obs}$ ratios, calculated using relation (\ref{L_ratio}).
The COMBO and low-redshift Seyfert 1 samples are both formed by low luminosity objects that we identify with Class S (Case A) objects, while the BQS sample with relatively high accretion rates can be identified with Class S (Case B) objects. The high-redshift AGNs and the QSO sample of \citet{J_et_2007} describe luminous and massive sources that we associate with Class Q objects.
Comparison of the observed values with the values predicted by our model therefore shows excellent agreement (see table 2). The main SDSS sample spans a wide range in luminosity and black hole mass illustrated by the intermediate $\left\langle \alpha_{OX} \right\rangle$ value. \\

\begin{table}    
\caption{Comparison of observational average $\left\langle L_{X}/L_{UV} \right\rangle_{obs}$ ratios of the different sub-samples with model $\left\langle L_{X}/L_{UV} \right\rangle_{model}$ ratios}
\label{table_obs}      
\centering                  
\begin{tabular}{l c c c}        
\hline\hline                 
Sample  (number of objects) & $\left\langle \alpha_{OX} \right\rangle$ &  $\left\langle L_{X}/L_{UV} \right\rangle_{obs}$ &  $\left\langle L_{X}/L_{UV} \right\rangle_{model}$\\    
\hline  
   low-redshift Seyfert 1  (37) & $-1.34^{(1)}$ & 0.39  & 0.40 $\dot{M}_{0}/M_{8}$ \\  
   COMBO  (47) & $-1.36^{(2)}$ & 0.34 & 0.40 $\dot{M}_{0}/M_{8}$ \\    
   BQS  (45) & $-1.46^{(2)}$ & 0.19 & 0.17\\             
   SDSS main  (155) & $-1.51^{(1)}$ & 0.14 & - \\      
   high-redshift luminous AGN  (36) & $-1.72^{(1)}$ & 0.04 & 0.01 $\dot{M}_{1}/M_{9}$  \\
   most luminous QSO  (33) & $-1.80^{(3)}$ & 0.02 & 0.01 $\dot{M}_{1}/M_{9}$\\
\hline 
                     
\end{tabular}
\begin{list}{}{}    
\item[$^{(1)}$] sample from \citet{S_et_2005}
\item[$^{(2)}$] sample from \citet{S_et_2006}
\item[$^{(3)}$] sample from \citet{J_et_2007}
\end{list}
\end{table}

   \begin{figure}
   \centering
   \includegraphics[width=0.5\textwidth]{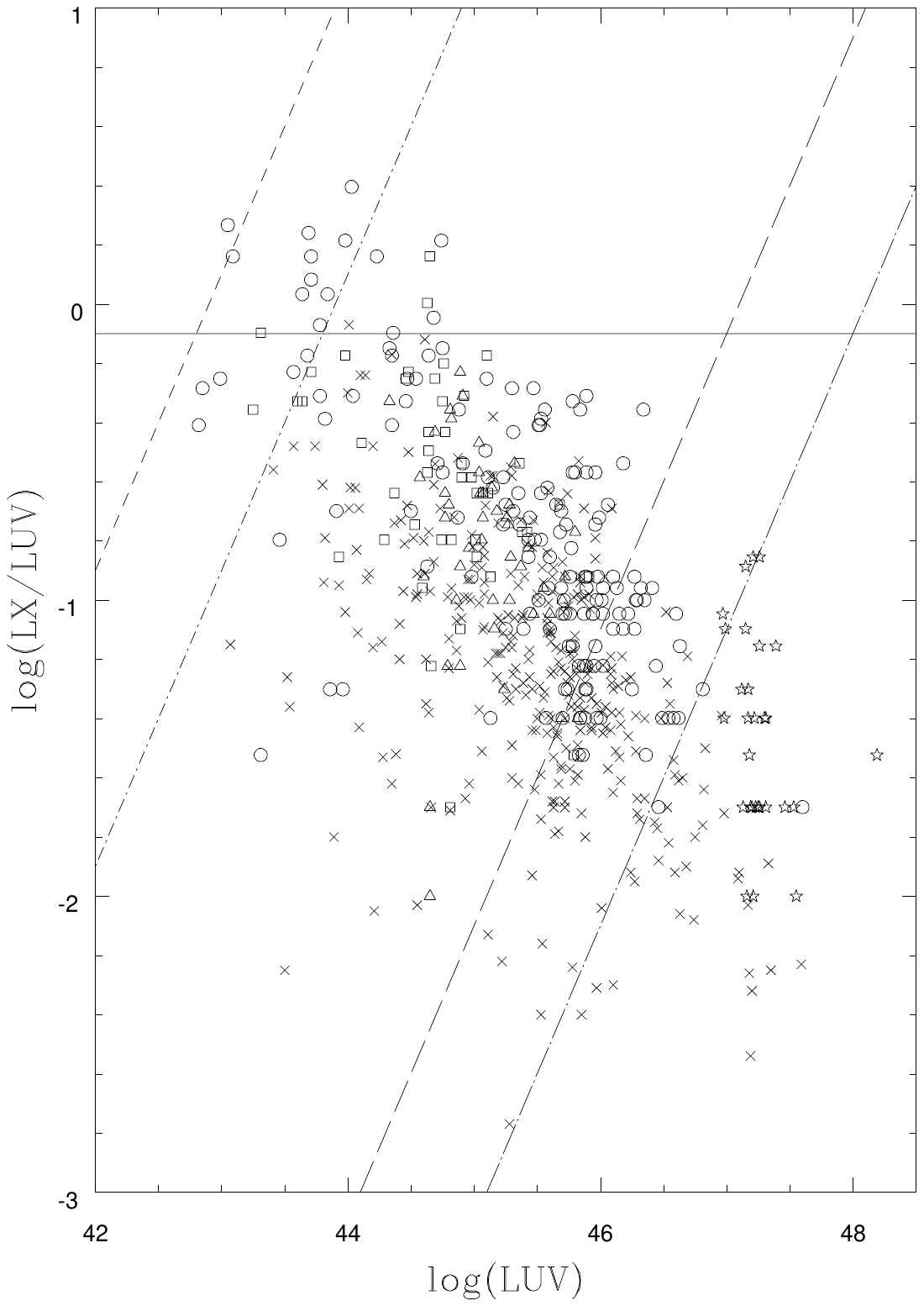}
      \caption{$log(L_{X}/L_{UV})$ versus $log(L_{UV})$ for the total sample. 
The main SDSS sample is represented by circles, the COMBO sample by squares, the BQS sample by  triangles, the luminous QSO sample by stars, and the \citet{K_et_2008} sample by crosses.
The upper diagonal lines represent Class S (Case A) relation with a central mass of $10^{7} M_{\odot}$ for $f_{1} = 1$ and $\eta_{1/3} = 0.3$ (short dash), $f_{1} = 0.5$ and $\eta_{1/3} = 1$ (dot-short dash). 
The solid horizontal line illustrates the upper limit for Class S (Case B), with $f_{1} = 1$ and $\eta_{1/3} = 0.3$. 
The lower diagonal lines show Class Q (Case A) relation with a central mass of $10^{10} M_{\odot}$ for $f_{1/2} = 2$ and $\eta_{1/3} = 0.3$ (long dash), $f_{1/2} = 1$ and $\eta_{1/3} = 1$ (dot-long dash).}
              
         \label{figure}
   \end{figure}

In figure 1, we plot $log(L_{X}/L_{UV})$ as a function of $log(L_{UV})$ for a sample spanning a large range in luminosity ($10^{43} \textrm{erg/s}  \lesssim L_{UV} \lesssim 10^{48} \textrm{erg/s}$). Model relations are shown for two values of the central mass ($10^{7} M_{\odot}$ and $10^{10} M_{\odot}$) with different shock parameters. The UV luminosities are directly given in the \citet{K_et_2008}'s sample as $log(L_{UV}/L_{E})$ where $L_{E}$ is calculated from the broad-line mass estimates; for the other samples, we have estimated the UV luminosity as $L_{UV} = \lambda L_{\lambda}$ at $\lambda = 2500 \AA$, from the given monochromatic luminositiy.

We see that the model predictions cover the right range of the observed properties, with the majority of objects lying within the expected values. We also note a trend of decreasing $L_{X}/L_{UV}$ ratios with increasing luminosity, which indicates that luminous, hence massive, objects tend to have lower $L_{X}/L_{UV}$ values compared to less massive sources. 
In our picture, larger $L_{X}/L_{UV}$ ratios are expected in Class S compared with Class Q. This result is thus in qualitative agreement with observations; it could explain the observed $L_{X}/L_{UV} - L_{UV}$ relation, considering that the distinction between quasars and Seyfert galaxies is mainly based on the central luminosity. Clearly, the most luminous sources (with $L_{UV} \gtrsim 10^{47} \textrm{erg/s}$) should have central masses exceeding $10^{10} M_{\odot}$. \\


\section{Discussion and Conclusion}

A thorough study of the relationship between different emission components is a required step towards a theoretical understanding of energy generation mechanisms in AGNs. 
It is hard to explain all the observed AGN properties within standard accretion disc models, the main difficulties including: the similarity in spectral features in the UV and X-ray domains observed in sources with huge differences in central luminosity, the origin of the X-ray emission and its relative importance compared to UV emission. \\
 
In the optical/UV domain, the spectral shape of the blue bump component is very similar in objects varying by 6 orders of magnitude in luminosity (\citet{W_F_1993}). In particular, the cut-off temperature of the bump is stable and does not vary by more than a factor of two in sources changing by a factor of $10^{4}$ in luminosity (\citet{W_et_1994}). 
However, standard accretion disc models do not predict such similarity in spectral features. The relatively universal value of the blue bump temperature on source luminosity is difficult to explain in the framework of standard discs in which the optical/UV emission explicitly depends on the black hole mass and accretion rate.
 
On the contrary, in the picture presented here, the optical/UV emission arises from optically thick shocks and the resulting blue bump temperature is independent of the black hole mass and only weakly dependent on collision parameters. 
Optically thick shocks would then naturally account for the observed similarity of the blue bump temperature in objects of very different luminosities, hence different central masses.
Our simple model may therefore explain the relatively universal value of the blue bump temperature in objects varying by several orders of magnitude in luminosity.   \\

In the X-ray domain, standard accretion discs cannot correctly account for the AGN emission. More complex models with additional components are therefore required, such as irradiated discs and disc-corona models. 
In the irradiated disc model, the disc emits as a result of both internal viscous heating and external radiative heating due to an X-ray point-like source located above it. But the origin and the location of the X-ray source irradiating the disc is a priori arbitrary and not physically justified. 
Moreover, in order to reproduce variations of similar amplitude in the UV and X-rays, the X-ray luminosity $L_{X}$ should be of the same order of the UV luminosity $L_{UV}$, which is contrary to  observations. 

A more physically plausible picture is given by the disc-corona model (\citet{H_M_1993}) in which the accretion disc is surrounded by a hot corona: the corona emits X-rays by Compton upscattering of the soft UV photons from the disc, while the disc reprocesses the X-ray photons from the corona into UV photons. 
But this model gives a larger $L_{X}/L_{UV}$ ratio than the observed value, as the corona and the disc luminosities are assumed to be of the same order. 
This led \citet{H_M_G_1994} to propose a variant consisting in a `patchy' corona, which leads to a decrease in the X/UV ratio.

An alternative model is given by the cloud model (\citet{C_et_1996}, \citet{C_D_1998}, \citet{A_C_2000}) in which a central X-ray source is surrounded by a number of Compton thick clouds in quasi-spherical geometry with a large coverage factor; this medium emits the blue bump and reprocesses the X-rays. Contrary to disc models, this latter model predicts a $L_{X}/L_{UV}$ ratio smaller than unity without any ad-hoc hypothesis due to the large coverage factor of the clouds. \\

In our picture, optically thick shocks give rise to optical/UV radiation, while the optically thin shocks are at the origin of the X-ray emission.
The production of X-rays does not require any additional component, since X-rays are emitted as a consequence of the Compton cooling process of electrons heated in the optically thin shocks.
The volume filling factor of the post-shock configuration plays an important role in determining the location of the optically thin shocks, and in defining two classes of objects, distinguished by the central mass and hence central luminosity, that we have identified with quasars and Seyfert galaxies. 
The competition between cooling and dynamical time scales suggests that there are two additional sub-classes for a given central mass, divided into high accretion rate and low accretion rate objects. 

Computing the ratio of X-ray luminosity to UV luminosity, $L_{X}/L_{UV}$, we obtain that this ratio is always smaller than unity, in agreement with the small X-ray to UV ratio observed in the majority of objects.
There are only few objects with $L_{X}/L_{UV}$ ratios exceeding unity in the total sample, and the majority of sources have luminosity ratios falling within the predicted range ($0.01 \lesssim L_{X}/L_{UV} \lesssim 0.8$). Our model is thus able to predict the observed range of $L_{X}/L_{UV}$ ratios, or equivalently the range of $\alpha_{OX}$ indices.
The observed $\alpha_{OX}-M_{BH}$ correlation in the sample of \citet{K_et_2008} implies that high mass objects have smaller $L_{X}/L_{UV}$ ratios, while low mass objects have larger $L_{X}/L_{UV}$ values. We also obtain here smaller $L_{X}/L_{UV}$ ratios in Class Q objects and larger $L_{X}/L_{UV}$ ratios in Class S objects, the predicted trend is thus consistent with the observational relation.
Our model may therefore suggest a possible explanation for the observed $\alpha_{OX}-l_{UV}$ anticorrelation. \\

Peculiar values of the $L_{X}/L_{UV}$ ratio, may be attributed to different factors such as additional X-ray emission from a jet component in radio-loud AGNs or absorption in BAL QSOs (considering residual sources not correctly removed in the sample selection). 
We should note that the observed $L_{X}/L_{UV}$ ratio in 3C 273 is one order of magnitude higher than the predicted value for massive objects. 
This discrepancy could be explained by an additional X-ray emission, probably associated with the jet component.\\

The existence of bulk relativisitc outflows in a preferred direction (collimated jets), cannot be accounted for within the proposed framework, as we have no privileged direction. 
However, the fraction of matter ejected outward following the expansion of the optically thick shock, may be associated with matter outflows observed in a number of AGNs. 
In order to obtain an outflow, the wind velocity should reach the local escape velocity and thus the launch radius should lie close to the escape radius. 
In our picture, the gas expansion velocity is given by the initial free-fall velocity (on the order of $\sim$ 0.1c at $100 \, \mathrm{R_{S}}$), which value coincides with the local escape velocity.
In addition, the mass outflow rate is comparable to the mass accretion rate, as indicated by observations. 
In the case of the QSO PG 1211+143, the outflow velocity is in the range 0.13-0.15c at an escape radius of $130 \, \mathrm{R_{S}}$ and with a mass outflow rate of $\dot{M}_{out} \sim 3.5 M_{\odot}/\mathrm{yr}$ (\citet{P_P_2006}). 
The location of the phenomenon and the outflow speed are compatible with shocks occuring at $\sim 100 \, \mathrm{R_{S}}$ and outflowing at the expansion velocity.
The expanding matter will be later slowed down in the outer regions, and may give rise to the line emitting clouds of the Broad Line Region.



\bibliographystyle{aa}
\bibliography{biblio}

\begin{thebibliography}{24}
\expandafter\ifx\csname natexlab\endcsname\relax\def\natexlab#1{#1}\fi

\bibitem[{{Abrassart} \& {Czerny}(2000)}]{A_C_2000}
{Abrassart}, A. \& {Czerny}, B. 2000, \aap, 356, 475

\bibitem[{{Collin-Souffrin}(1991)}]{C_1991}
{Collin-Souffrin}, S. 1991, \aap, 249, 344

\bibitem[{{Collin-Souffrin} {et~al.}(1996){Collin-Souffrin}, {Czerny},
  {Dumont}, \& {Zycki}}]{C_et_1996}
{Collin-Souffrin}, S., {Czerny}, B., {Dumont}, A.-M., \& {Zycki}, P.~T. 1996,
  \aap, 314, 393

\bibitem[{{Courvoisier}(2001)}]{C_2001}
{Courvoisier}, T.~J.-L. 2001, in Quasars, AGNs and Related Research Across
  2000. Conference on the occasion of L. Woltjer's 70th birthday, ed.
  G.~{Setti} \& J.-P. {Swings}, 155--+

\bibitem[{{Courvoisier} \& {Clavel}(1991)}]{C_C_1991}
{Courvoisier}, T.~J.-L. \& {Clavel}, J. 1991, \aap, 248, 389

\bibitem[{{Courvoisier} \& {T{\"u}rler}(2005)}]{C_T_2005}
{Courvoisier}, T.~J.-L. \& {T{\"u}rler}, M. 2005, \aap, 444, 417

\bibitem[{{Czerny} \& {Dumont}(1998)}]{C_D_1998}
{Czerny}, B. \& {Dumont}, A.-M. 1998, \aap, 338, 386

\bibitem[{{Falk} \& {Arnett}(1977)}]{F_A_1977}
{Falk}, S.~W. \& {Arnett}, W.~D. 1977, \apjs, 33, 515

\bibitem[{{Gondek} {et~al.}(1996){Gondek}, {Zdziarski}, {Johnson}, {George},
  {McNaron-Brown}, {Magdziarz}, {Smith}, \& {Gruber}}]{G_et_1996}
{Gondek}, D., {Zdziarski}, A.~A., {Johnson}, W.~N., {et~al.} 1996, \mnras, 282,
  646

\bibitem[{{Haardt} \& {Maraschi}(1993)}]{H_M_1993}
{Haardt}, F. \& {Maraschi}, L. 1993, \apj, 413, 507

\bibitem[{{Haardt} {et~al.}(1994){Haardt}, {Maraschi}, \&
  {Ghisellini}}]{H_M_G_1994}
{Haardt}, F., {Maraschi}, L., \& {Ghisellini}, G. 1994, \apjl, 432, L95

\bibitem[{{Just} {et~al.}(2007){Just}, {Brandt}, {Shemmer}, {Steffen},
  {Schneider}, {Chartas}, \& {Garmire}}]{J_et_2007}
{Just}, D.~W., {Brandt}, W.~N., {Shemmer}, O., {et~al.} 2007, \apj, 665, 1004

\bibitem[{{Kelly} {et~al.}(2008){Kelly}, {Bechtold}, {Trump}, {Vestergaard}, \&
  {Siemiginowska}}]{K_et_2008}
{Kelly}, B.~C., {Bechtold}, J., {Trump}, J.~R., {Vestergaard}, M., \&
  {Siemiginowska}, A. 2008, ArXiv e-prints, 801

\bibitem[{{Koratkar} \& {Blaes}(1999)}]{K_B_1999}
{Koratkar}, A. \& {Blaes}, O. 1999, \pasp, 111, 1

\bibitem[{{Malkan}(1983)}]{M_1983}
{Malkan}, M.~A. 1983, \apj, 268, 582

\bibitem[{{Paltani} {et~al.}(1998){Paltani}, {Courvoisier}, \&
  {Walter}}]{P_C_W_1998}
{Paltani}, S., {Courvoisier}, T.~J.-L., \& {Walter}, R. 1998, \aap, 340, 47

\bibitem[{{Pounds} \& {Page}(2006)}]{P_P_2006}
{Pounds}, K.~A. \& {Page}, K.~L. 2006, \mnras, 372, 1275

\bibitem[{{Shakura} \& {Syunyaev}(1973)}]{S_S_1973}
{Shakura}, N.~I. \& {Syunyaev}, R.~A. 1973, \aap, 24, 337

\bibitem[{{Shields}(1978)}]{S_1978}
{Shields}, G.~A. 1978, \nat, 272, 706

\bibitem[{{Soldi} {et~al.}(2008){Soldi}, {T{\"u}rler}, {Paltani}, {Aller},
  {Aller}, {Burki}, {Chernyakova}, {L{\"a}hteenm{\"a}ki}, {McHardy}, {Robson},
  {Staubert}, {Tornikoski}, {Walter}, \& {Courvoisier}}]{S_et_2008}
{Soldi}, S., {T{\"u}rler}, M., {Paltani}, S., {et~al.} 2008, \aap, 486, 411

\bibitem[{{Steffen} {et~al.}(2006){Steffen}, {Strateva}, {Brandt}, {Alexander},
  {Koekemoer}, {Lehmer}, {Schneider}, \& {Vignali}}]{S_et_2006}
{Steffen}, A.~T., {Strateva}, I., {Brandt}, W.~N., {et~al.} 2006, \aj, 131,
  2826

\bibitem[{{Strateva} {et~al.}(2005){Strateva}, {Brandt}, {Schneider}, {Vanden
  Berk}, \& {Vignali}}]{S_et_2005}
{Strateva}, I.~V., {Brandt}, W.~N., {Schneider}, D.~P., {Vanden Berk}, D.~G.,
  \& {Vignali}, C. 2005, \aj, 130, 387

\bibitem[{{Walter} \& {Fink}(1993)}]{W_F_1993}
{Walter}, R. \& {Fink}, H.~H. 1993, \aap, 274, 105

\bibitem[{{Walter} {et~al.}(1994){Walter}, {Orr}, {Courvoisier}, {Fink},
  {Makino}, {Otani}, \& {Wamsteker}}]{W_et_1994}
{Walter}, R., {Orr}, A., {Courvoisier}, T.~J.-L., {et~al.} 1994, \aap, 285, 119

\end{thebibliography}

\end{document}